\input psfig
\documentstyle[twocolumn,aps]{revtex}
\begin{document}
\draft

\twocolumn[\hsize\textwidth\columnwidth\hsize\csname @twocolumnfalse\endcsname

\title{ Rapid Suppression of the Spin Gap in Zn-doped
CuGeO$_3$ and SrCu$_2$O$_3$ }

\author{George Balster Martins and Elbio Dagotto}
\address{National High Magnetic Field Lab and 
Department of Physics, Florida State University, Tallahassee,\\
Florida 32306,USA}
\author{Jos\'{e} A. Riera}
\address{Instituto de Fisica Rosario, Av. 27 de Febrero 210 bis, 2000 Rosario,\\
Argentina}
\maketitle

\begin{abstract}
The influence of non-magnetic impurities on the spectrum and dynamical
spin structure factor of a model for CuGeO$_3$ is studied. A simple
extension to Zn-doped ${\rm Sr Cu_2 O_3}$ is also discussed.
Using Exact
Diagonalization techniques and intuitive arguments we show that
Zn-doping introduces states in the Spin-Peierls gap of CuGeO$_3$. This effect can be
understood easily in the large dimerization limit where doping by Zn
creates  ``loose'' S=1/2 spins, which interact with each other through very weak effective
antiferromagnetic couplings. When the dimerization is small, a similar
effect is observed but now with the free S=1/2 spins being the resulting S=1/2 ground
state of severed chains with an odd number of sites. Experimental
consequences
of these results are discussed. It is interesting to observe that the
spin correlations along the chains are enhanced by Zn-doping according
to the numerical data presented here.
As recent numerical calculations have shown, 
similar arguments apply to ladders with non-magnetic impurities simply
replacing the tendency to dimerization in CuGeO$_3$ by
the tendency to form spin-singlets along the rungs in SrCu$_2$O$_3$.

\end{abstract}


\vskip2pc]
\narrowtext

\section{introduction}

The recent discovery of a spin-Peierls (SP) transition in the
inorganic compound ${\rm CuGeO_3}$
has received considerable attention.\cite{hase1} The effect 
is presumably caused by the coupling of the spins along 
the ${\rm Cu-O_2}$ chains with three-dimensional phonons. This coupling
dimerizes the chains reducing (increasing) the lattice spacing 
for, e.g., the even (odd) links, and thus promoting the formation of spin singlets
along the short bonds. A spin-gap is formed in the spectrum, which is roughly
the energy necessary to break the even-bond spin singlets. Soon after the
discovery of this material, studies where ${\rm Cu^{2+}}$ (S=1/2) was 
substituted by ${\rm Zn^{2+}}$, ${\rm Ni^{2+}}$ or
${\rm Mn^{2+}}$ were reported.\cite{hase2} Based on 
magnetic susceptibility $\chi$ 
data analysis, a rapid suppression of the SP spin-gap was
observed, with a small 2\% concentration of Zn being sufficient to induce
a transition to a new
phase. $\chi$, specific heat and inelastic neutron scattering (INS)
 measurements\cite{oseroff,hase3,lussier}
have unambiguously
shown that this new phase is a three dimensional (3D) N\'eel
state. The simultaneous existence in the experimental data 
of features 
indicating SP and N\'eel order has also been noticed.\cite{oseroff}
Similar results have been reported with Si substituting Ge,\cite{renard}
and Ni,Mn,Mg replacing Cu.\cite{oseroff,ajiro}
These results are apparently inconsistent with the gapless SP state
predicted by mean-field theory.\cite{lu}
Recently\cite{sasago}, INS
measurements in good quality
samples   of the temperature dependence of the
superlattice peaks caused by  the lattice dimerization 
were interpreted as
evidence that the SP actually survives up to 6\% Zn-doping.

The experimental evidence accumulated recently 
in doped ${\rm CuGeO_3}$ shows two
interesting phenomena 
namely (i) the unexpectedly rapid suppression of 
the SP order by Zn-doping, and (ii) its replacement by 3D AF order.
Since INS 
results for 2\% Zn-doping\cite{sasago} show already an important reduction 
of the SP transition at temperatures well above those where AF order starts,
we conjecture that these two phenomena could be studied  independently.
It is remarkable that similar experimental 
results\cite{azuma} have been observed in a
quasi one-dimensional ``ladder'' 
system, ${\rm Sr (Cu_{1-x} Zn_x )_2 O_3}$, where 
a spin-gap exists at x=0 due to the formation of spin-singlets along
the ladder rungs,\cite{ladder} without lattice deformations. 
This suggests that the rapid suppression
of a spin-gap by Zn-doping and its subsequent replacement by 3D N\'eel order
may be a general phenomenon independent of the
origin of such a  spin-gap.

In this paper, the effect of non-magnetic Zn-doping on the 
SP state of ${\rm CuGeO_3}$ is studied. A simple generalization to the
case of Zn-doped ladders is also discussed.
Using numerical techniques we show
that the SP spin-gap is rapidly suppressed with Zn-doping, as observed
experimentally. The effect is caused by unpaired spins created by the
breaking of spin-singlets upon doping (in agreement with conclusions
reached in recent numerical studies of even-leg ladder systems\cite{sandvik}). 
These unpaired spins coupled 
forming an effective random spin-1/2 chain.
Our analysis does not include
intra-chain couplings that could stabilize the 3D AF order, analysis that
is postponed for a future publication\cite{fukuyama} (however, in Sec.
VII below we discuss the enhancement observed numerically 
of spin correlations upon Zn doping. This effect may contribute to the
stabilization of the 3D AF order in real materials).
To study the dimerization of a 
spin-1/2 chain it is natural to use the standard Heisenberg model
including a static modulation of the nearest-neighbor (NN) exchange to
account for the lattice distortion. In addition, recent studies have shown
that a next-nearest-neighbor (NNN) spin-spin interaction is also needed
to properly fit experimental susceptibility data.\cite{riera1,castilla,haas,riera2}
This is reasonable based on the structure of ${\rm CuGeO_3}$ where
Cu-Cu interactions bridged by two oxygens, Cu-O-O-Cu, may not be negligible.  
Thus, the model used here is
$$
{\cal H}  =  J_1 \sum_{\langle ij \rangle } {{\bf
S}_i\cdot {\bf S}_j} +J_2 \sum_{\langle \langle
ik \rangle \rangle }{ {\bf S}_i\cdot {\bf S}_k}
-
$$
$$
J_1\delta \sum_{\langle ij \rangle } ( -1)^i
{{\bf S}_i\cdot {\bf S}_j },
\eqno(1)
$$
where $\langle ij \rangle$ ($\langle \langle ik \rangle \rangle$) denote NN
(NNN) sites, $\delta$ regulates the amount of dimerization, and $J_1, J_2$
are the NN and NNN Heisenberg couplings,  respectively. The rest of the
notation is standard. A modulation of $J_2$ is a higher order effect neglected
here. To model ${\rm CuGeO_3}$, we here use $\delta=0.03$ and 
$J_2/J_1 = 0.24$,\cite{castilla}
but other set of parameters could be chosen.\cite{riera1} The qualitative
results described 
below do not depend crucially on the actual values of $\delta$ and 
$J_2/J_1$.
The introduction of a non-magnetic impurity at site $i$ is mimicked in Eq.(1)
by removing the $J_1$, $J_2$ and $\delta$-terms associated with this site.
Note that the $J_2$ NNN coupling between 
sites $i-1$ and $i+1$ is not influenced
by an eventual impurity located at $i$. This coupling across impurities
 plays an important role in 
linking chain segments severed by Zn-doping.

\section{Large Dimerization Limit}

To guide the intuition on how  Zn-doping affects the 
properties of model Eq.(1), 
it is convenient to study the limit
where the dimerization $\delta$ is large, inducing a 
well-defined pattern of dimers (spin-singlets) in the ground state of the system.
To illustrate the physics of the problem
consider in Fig.1 a simple case where
two impurities are randomly distributed on an 8-site chain using periodic
boundary conditions (PBC). Impurity-1 is located at site 1, and the second one
at different sites along the chain. There are six independent cases
shown in
Fig.1.
In Fig.2a, the longitudinal dynamical
spin structure factor $S^{zz}(\pi ,\omega )$, which is the spectral 
decomposition of the operator $(1/N)
\sum_i (-1)^i S^{z}_i$, is shown for the different  positions of the
impurities ($N$ is the number of sites). We consider momentum $\pi$
because in this subspace is where the smallest gap is found in the
undoped system.
The calculation was performed with standard
Exact Diagonalization (ED) techniques,\cite{review}
and using $\delta=0.5$ (providing a strong exchange $J=1.5J_1$ for half the
links of the chain, 
and a weak exchange $J=0.5J_1$ for the other half). We observed that for
 such a large dimerization finite size
effects are small. We also verified that using chains with open 
boundary conditions (OBC) the conclusions of this
discussion are unchanged.

\begin{figure}[htbp]
\centerline{\psfig{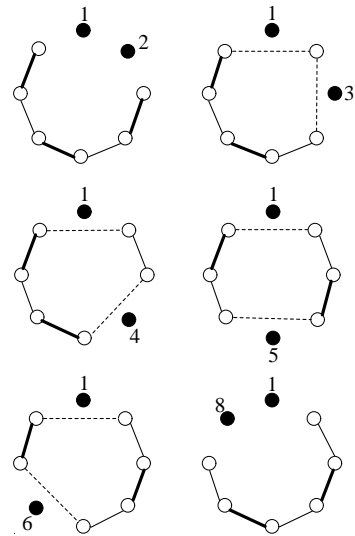}}
\vspace{0.5cm}
\caption{Eight site chain with PBC and two impurities located in the
positions of the full dots. The thick (thin) 
links denote a large (small) spin-spin
coupling of magnitude $J_1(1+\delta)$ ($J_1(1-\delta)$). The dashed line
linking some pairs of sites illustrates the presence of the $J_2$ coupling
across impurities. The other $J_2$ couplings are not shown explicitly.
The case of impurities at sites 1 and 7 is identical to having impurities
at 1 and 3.}
\end{figure}

Let us consider the six impurity distributions depicted in Fig.1: (1) 
for impurities at sites 1-2 a strong dimer is entirely removed. All the
other spins are themselves  strongly coupled in dimers, and thus we expect
spin excitations at energies $\omega \sim 1.5 J_1$ which are 
needed to transform a
spin singlet into a triplet, in good agreement with Fig.2a. In all the 
other cases below, weight at this large $\omega$ is also expected due to 
the presence of strong dimers in the ground state; (2) if the 
impurities are at sites 1 and 3, they break two dimers producing two $unpaired$ spins.
These spins tend to form a spin-singlet using the coupling $J_2$, since the
rest of the spins are mostly forming dimers as shown in Fig.1. Then, singlet-triplet excitations
are expected at $\omega \sim 0.24J_1$. The anomalous small weight of the
associated pole in Fig.2a is caused by the use of momentum $\pi$ in the
dynamical spin structure factor that gives small weight to singlets
where both spins are at even or odd sites; 
(3) with impurities at 1 and 4, two strong dimers
are erased, and the two remaining spins are linked into a singlet
through $J=0.5J_1$
inducing a pole at energy $\omega \sim 0.5J_1$ as in Fig.2a; (4) when the
impurities are at opposite sites of the chain, two free spins are created which
are linked by a very weak effective exchange
since two of the intermediate spins 
form a strong singlet. This would imply the presence of weight at small $\omega$.
However, sites 2 and 6 are both even and the spin structure factor at
$q=\pi$ suppresses the signal; (5) with impurities at sites 1-6, the two
free spins are also weakly coupled into a singlet (weight at small $\omega$ is
thus expected in agreement with Fig.2a),
but now being at even and odd sites their
contribution to $S^{zz}(\pi ,\omega )$ is not suppressed by the sign modulation.
The effective coupling is $J_{eff} \sim 0.07 J_1$;
(6) finally, for impurities at 1 and 8, the free spins are again weakly linked
and the intensity of the pole signaling their singlet-triplet
excitation
 is enhanced by the momentum $q=\pi$. Thus, weight is expected
at small $\omega$, as in Fig.2a. Note that in this case the effective exchange is the smallest
of the cases shown in
Fig.1, $J_{eff} \sim 0.02 J_1$, since it is mediated by two strong dimers. 

\begin{figure}[htbp]
\centerline{\psfig{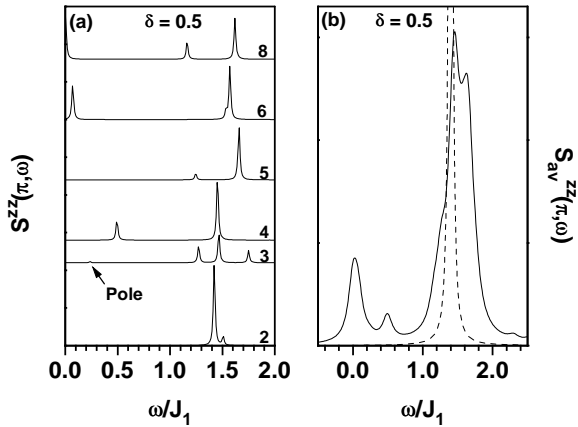}}
\vspace{0.5cm}
\caption{(a) $S^{zz}(\pi,\omega)$ vs $\omega$ on an eight site chain with
PBC and using a dimerization $\delta = 0.5$. 
The impurities are distributed as in Fig.1. 
The $\delta$-functions have been
given a width $\epsilon = 0.03$. 
The labels indicate
the position of the second impurity, with the first one 
located at site 1; (b) $S_{av}^{zz}(\pi,\omega)$ 
(average of results Fig.2a over the positions of the impurities) 
vs $\omega$ (solid line).
The dashed line indicates the result in the absence of impurities.}
\label{fig2}
\end{figure}

In Fig.2b, the equal-weight average of $S^{zz}(\pi,\omega )$
over the impurity positions, denoted as  $S_{av}^{zz}(\pi ,\omega )$,
 is shown. The result  in the absence of
impurities is also presented. It is clear that the introduction of impurities
broadens the main peak, and, more importantly, it introduces weight inside
the original spin-gap. The peak the smallest in intensity  is caused by 
the special case when two impurities are at close distance, while the weight
at very small $\omega$ is originated by the weak coupling into singlets
of unpaired spins at large distances. The weight inside
the gap should diminish as the density of impurities is reduced.
Note that the effective interaction  between spins at large distances
can be understood using 
a decimation procedure analog to that employed by 
Dasgupta and Ma\cite{ma}, in their study of random spin-1/2 chains.

\section{realistic parameters}

In Fig.3a, $S_{av}^{zz}(\pi ,\omega )$ is shown as 
$\delta $ varies from 0.5 to 0.03 now using a chain 
of 20 sites and still 2 impurities. The peak at high $\omega$ moves down
in energy as $\delta$ is reduced, as expected since it
is originated by the
singlet-triplet excitations of the dimers. On the other hand, 
the weight at small $\omega$
increases slightly its energy as $\delta$ is reduced. This is also compatible with
the expectation that this feature is created by  the AF coupling 
between ``free'' spins located at large distances 
from each other. A smaller $\delta$ increases the coupling
between those spins since the tendency to form tight intermediate spin
singlets is reduced. The evolution with $\delta$ of Fig.3a is smooth and
suggests that the intuitive explanation for the filling of the gap found
at large $\delta$ could be qualitatively correct even in the more realistic
regime of $\delta \ll 1$.

\begin{figure}[htbp]
\centerline{\psfig{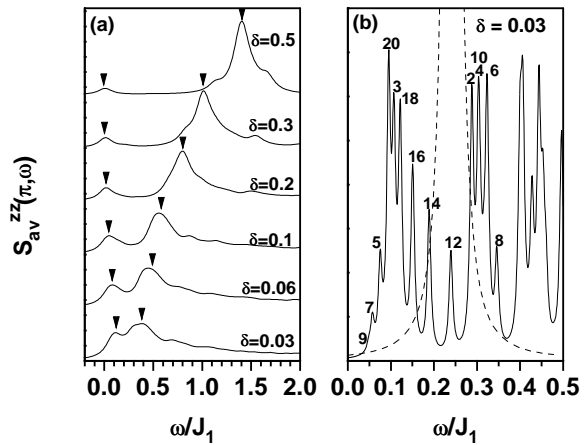}}
\vspace{0.5cm}
\caption{(a) $S_{av}^{zz}(\pi,\omega)$ vs $\omega$ on a 20-site chain 
parametric with $\delta$ as indicated. The peaks in the spectral weight
have been given a width $\epsilon = 0.03$.
(b) $S_{av}^{zz}(\pi,\omega)$ vs $\omega$ for $\delta = 0.03$ with 
a resolution $\epsilon = 0.006$. The labels of the peaks refer to the
position of a second impurity on a chain of 20-sites with the first impurity
located at site 1. The missing sites are related to those shown by
symmetry, or they have negligible weight.}
\label{fig3}
\end{figure}

On Fig.3b, $S_{av}^{zz}(\pi ,\omega )$ is shown for $\delta
=0.03 $ but with a higher resolution than used in Fig.3a. 
Each peak represents the position of an excited state of the system
with spin 1. The labels refer to the position of the second impurity,
the first one being always at site 1. With this resolution, we can
identify the origin of the two dominant features observed in the
spin structure factor Fig.3a at small $\delta$,
namely the presence of low-$\omega$  and high-$\omega$ 
spectral weight. Fig.3b shows a clear
difference between even and odd sites. To understand this effect, note
that the case where the 
second impurity is on an odd-site corresponds to a partition of the
chain into two segments each with an $odd$ number of spins. This
effectively produces two S=1/2 states in each segment which are coupled
by the relatively small $J_2$. The fact that such an effective ``free''
 spin-1/2 in each segment is now spread over more than one lattice
spacing contributes to the reduction of the effective exchange leading
to the very small energy excitation of, e.g., the case where the 
impurities are at sites 1 and 9.
Thus, the same reasoning leading to the appearance of weight at small
$\omega$
in the large $\delta$ limit, can be repeated for small $\delta$ still
using $J_2$ as the link between S=1/2 effective spins. 

\section{intuitive picture}

The basic idea found numerically in the previous section
is that Zn-doping divides a S=1/2 chain into segments each carrying
either a total spin 1/2 or 0, depending on the number of spins in the segment.
Then, at low temperatures each segment can be replaced by just one
effective  spin
with coupling that depend on the distance between impurities (i.e. on
the length of the segments).
This is reminiscent of a similar approach followed 
recently to study  spin-1/2 chains with randomly distributed
ferromagnetic and antiferromagnetic bonds, as realized in
${\rm Sr_3 Cu Pt_{0.5} Ir_{0.5} O_6}$. \cite{furusaki}
If such a simple picture were correct for Zn-doped ${\rm Cu Ge O_3}$, then
the low-energy states of the complete doped chain could be
calculated using an effective Heisenberg Hamiltonian that couples
the effective neighboring spins.

\begin{figure}[htbp]
\centerline{\psfig{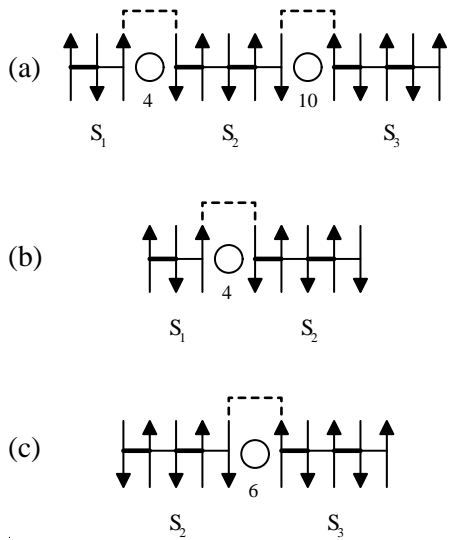}}
\vspace{0.5cm}
\caption{(a,b,c) Clusters used to illustrate the main intuitive ideas
explained in the text. Each of the three segments carry an effective
spin 1/2, denoted by ${\bf S_1}$, ${\bf S_2}$ and ${\bf S_3}$. The
circles
are the impurities. The thick (thin) bonds have large (small)  exchanges.
The dashed line denotes some of the $J_2$ couplings.}
\label{fig4}
\end{figure}

Let us verify this scenario with the example of a 15-sites chain with OBC, and two
impurities located at sites 4 and 10 dividing the lattice into
three segments, each with an odd number of spins (Fig.4a).
This setup is supposed to produce 
three effective spin-1/2 that we denote as $S_1$, $S_2$
and $S_3$. Its low energy Hamiltonian should be
$$
{\cal H} =  J_{12} {{\bf S}_1\cdot {\bf S}_2} + J_{23}
{{\bf S}_2\cdot {\bf S}_3},
\eqno(2)
$$
with $J_{12}$ and $J_{23}$ denoting effective exchange couplings, which
in principle correspond to the singlet-triplet energy separation $\Delta_{ST}$
for the case where two adjacent segments of Fig.4a are analyzed in
isolation. Diagonalizing exactly these coupled segments of 3 and 5
sites (Fig.4b) and 5 and 5 sites (Fig.4c), respectively, $J_{12}$ and
$J_{23}$ have been calculated, and the resulting couplings used in the
full three-segment 15-site problem of Fig.4a, which can also be exactly
diagonalized to gauge the accuracy of Eq.(2). Fig.5 shows that the
agreement between the exact result and the effective model is excellent
providing support to the simple picture outlined above.

\begin{figure}[htbp]
\centerline{\psfig{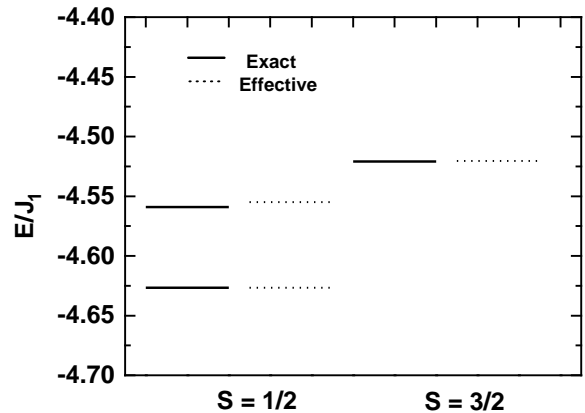}}
\vspace{0.5cm}
\caption{Energy levels of the effective Hamiltonian Eq.(2) compared against
those
of the 15-site cluster shown in Fig.4a. The eigenvalues of Eq.(2) 
were shifted by a constant
such that its ground state energy coincides with that of the 15-site chain.}
\label{fig5}
\end{figure}

To complete this analysis, we must investigate the case where the
segments have an even number of spins. Here in the large $\delta$ limit
there are two clearly different situations: 1. all the spins of the
even-site segment are coupled into strong dimers or 2. there are two ``loose'' spins at
the ends of the segment. In the first case, it is clear that an even-site
mostly dimerized segment  intercalated between two odd-spin chains
simply reduces substantially the coupling between the effective spins of
such odd-chains (with the effect becoming more drastic as the length of
the even-chain grows). This implies that the random distribution of
antiferromagnetic couplings resulting from this effect may contain large
weight at small exchanges i.e. it will not be centered at $J_1$ or $J_2$.
In the second case where there are two loose spins in
the same segment, their mutual interaction is also weak since the rest
of the spins form mostly singlets at least at large $\delta$. However, the free
spins at the ends of the segment can couple through $J_2$ with the free
spins of the adjacent chains.
We have carried out explicit ED calculations that indeed confirm this simple picture.

\section{comparison with available experiments}

In Fig.6a, the dynamical spin structure factor
$S_{av}^{zz}(q,\omega )$, corresponding to the spectral decomposition of
the operator $(1/N)
\sum_l e^{iql} S^{z}_l$,  is shown parametric with momentum $q$. 
As discussed before, this is an ``equal-weight'' average over the
positions of the impurities. Note that in real materials we may have some other
probability distribution of hole positions determined by energy and
entropy factors. Keeping this detail in mind, we proceed with the
analysis of Fig.6a assuming an equal-weight distribution.
For all
momenta there is a clear distinction between the low and high-$\omega$
features discussed before. Fig.6b contains the $q$-dependence of the
centroid of the low and high-$\omega$ spectral weight. It is clear that the
features at large $\omega$ are remnants of the S=1 excitations of the
$J_1$-$J_2$ model (here $\delta=0.03$ plays only a secondary role).\cite{haas} 
These results also clearly show that weight is generated for energies inside
the original SP spin-gap of the undoped system upon non-magnetic doping of
the chains. This weight has a mild $q$-dependence.
Thus, based on the intuitive picture developed in this paper
the spin-gap is, rigorously speaking,
destroyed for an infinitesimal concentration of Zn-doping. However,
the weight inside the gap is also proportional to the impurity
concentration and it may not be experimentally detectable until a finite
doping threshold is reached.

\begin{figure}[htbp]
\centerline{\psfig{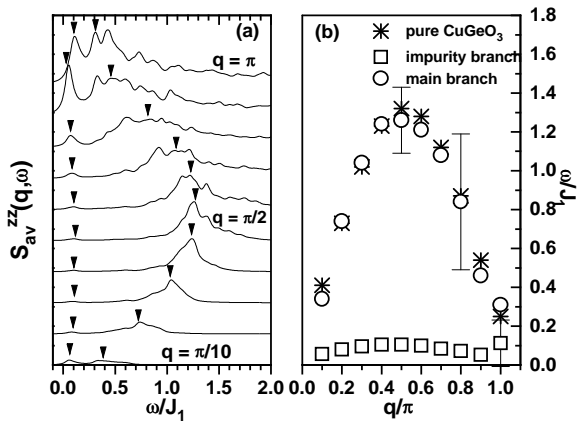}}
\vspace{0.5cm}
\caption{(a) $S_{av}^{zz}(q,\omega )$ vs $\omega$ for a 20-site chain with 2
impurities, parametric with $q$. The triangle denotes the rough position
of the low and high-$\omega$ spectral weight; (b) Position of the dominant high
and low-$\omega$ features of (a) compared with results in the absence of impurities.}
\label{fig6}
\end{figure}

It is important to remark that 
even though the gap is
closed with doping, in Fig.6a clear remnants of the original undoped system
are observed. Then, according to our picture  INS
experiments should indicate the presence of dimerization and AF
correlations for Zn-concentrations where other probes, specially
magnetic susceptibility measurements, indicate the disappearance of the spin-gap.
Short-distance dimerization correlations tested by INS
can survive the introduction of
doping up to large impurity concentrations.
However, $\chi$ defined as
$$
\chi  \propto lim_{q \rightarrow 0}  \int_0^\infty 
d\omega \frac{S^{zz}(q,\omega )}\omega
\eqno(3)
$$
is dominated by small energy excitations even if they carry small
weight. Using $q=\pi/10$ in Eq.(3), we calculated numerically  $\chi$
at zero temperature
using a 20-site chain with 0 and 2 impurities. The ratio
$\chi_{doped}/\chi_{undoped}$ is about 40.
Therefore this explains why $\chi$ measurements can show no trace of a
SP phase, while INS still suggest its presence.\cite{sasago} It is
the  short- against long-character of the correlations that establish the
difference. $\chi$ reacts to the long distance structure, while INS
captures fluctuations at all distances.

\section{experimental predictions}

   The numerical and analytical results discussed in this paper suggest that
${\rm CuGeO_3}$ looses its gap immediately
after an infinitesimal amount of Zn replaces Cu. 
This dramatic effect produces
interesting experimental consequences in addition to those described
in the previous section.
The localized spin-1/2 states in the vicinity of the nonmagnetic ions
interact weakly  through a Heisenberg
Hamiltonian with antiferromagnetic couplings having a strength that
depends on the distance between the Zn impurities. Since this distance
is random, the couplings are random and locate the low energy behavior
of Zn-doped ${\rm CuGeO_3}$ in the universality class of the random
exchange Heisenberg model.
This low energy effective model, defined by the Hamiltonian
$$
H = \sum_{\langle ij \rangle } J_{ij} {{{\bf S}_i}\cdot{{\bf S}_j}}
$$
in a standard notation, has been analyzed using decimation 
techniques.\cite{ma,fisher,boechat}
In this method the pair of spins with strongest coupling in  the random
chain is eliminated by considering the interaction with the neighboring
pairs as a perturbation. With this procedure a new (small) 
coupling is generated
between the two spins neighboring to the eliminated pair. The iteration
continues until a single pair of spins remains. As this decimation procedure
evolves, the original initial distribution $P_0(K)$ of random bonds
($0 < K < J$)
changes into
a power-law distribution $P(K) \sim K^{\alpha -1}$, where $\alpha$ 
depends only weakly with the "cutoff" $J$. In this asymptotic regime,
corresponding experimentally  to sufficiently low temperatures,
the susceptibility behaves as $\chi \sim T^{-\alpha}$, and
the magnetic contribution to the
specific heat as $C \sim T^{1-\alpha}$ 
(``random singlet phase''). Since $\alpha$ is positive, the
susceptibility
diverges as the temperature approaches zero.
For magnetic fields much stronger
than the temperature but much weaker than $J$ (all in energy units),
the magnetization behaves as $m \sim H^{1-\alpha}$
(note that
the actual value of $\alpha$ is difficult to
calculate within the accuracy of our techniques since we need to know
the initial distribution of random exchange couplings).
These predictions have been verified in 
organic materials and also in ${\rm MgTiOBO_3 }$\cite{fernandez}
where it was found that $\alpha \sim 0.8$. Thus, if indeed the low energy
behavior of Zn-doped ${\rm CuGeO_3}$ can be mapped into a random 
Heisenberg model, just one parameter ($\alpha$) is needed to fit
the power-law behavior of the susceptibility, specific heat and magnetization
as described above. 
This is a  prediction of our theory
for Zn-doped spin-Peierls systems in the regime of small Zn
concentration, and it breaks down at higher concentrations
when the experimentally observed transition to a long-range 3D N\'eel order
takes place.

\section{conclusions, extension to ladders, and enhancement of spin correlations}

Summarizing, here we have developed a simple picture of the effect of
Zn-doping on compounds with a spin gapped ground state. The idea is
that each non-magnetic impurity produces an effective S=1/2 ``free'' spin
which may be an actual S=1/2 electron or hole located at a 
nearest-neighbor site (in the limit of large dimerization),
or it may correspond to many-body S=1/2 states of severed odd-site
chains (in the limit of weak dimerization). Similar concepts have been
used
in the context of other compounds that provide physical realizations of
random ferro-antiferromagnetic exchange S=1/2 chains.\cite{furusaki}
The interaction of the free spins is mediated by the rest of the spins,
and thus it can be very small at low Zn-concentration. Such a weak
coupling induces low energy spectral weight inside the spin gap in 
calculations of the dynamical spin structure factor. This effect occurs
at all Zn densities, with a weight proportional to such density.
Clear remnants of 
the dynamical features corresponding to an undoped chain are observed.
 On the other hand,
the magnetic susceptibility, which reacts to low energy excitations,
indicates the rapid suppression of the spin gap at very low impurity
concentrations. Thus, an analysis of INS and $\chi$ data 
may naively produce different
conclusions unless we recall that INS reacts to 
fluctuations at all distances, while $\chi$ is 
affected mostly by the long-range order in
the system.

The calculations presented in this paper have been carried out for the
special case of Zn-doped CuGeO$_3$, but the results can be easily
generalized to other system with spin-gaps.
Of particular interest are the recent results reported
in ladders,\cite{azuma} where a rapid collapse of the spin-gap was observed.
To understand this result
consider the limit where the rung exchange $J_{\perp}$
is dominant (which is the analog of the large $\delta$ limit 
studied in Sec.II), and thus
the ground state is made out of rung spin-singlets.\cite{ladder} In this case
each Cu replaced by Zn actually induces  a loose S=1/2 spin,
as recently observed numerically by Sandvik et al.\cite{sandvik}
(see Fig.7).
 
\begin{figure}[htbp]
\centerline{\psfig{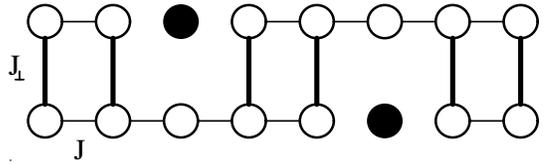}}
\vspace{0.5cm}
\caption{Schematic representation of a Zn-doped ladder. The open circles are
spins and the full circles are impurities. The thick bonds are the
strongest in the large $J_{\perp}$ limit.}
\label{fig7}
\end{figure}

Their interaction should lead to small spectral weight in the dynamical 
structure factor, similar to the results for the SP ground
state. Then, INS studies of Zn-doped
ladders at a small Zn concentration would indicate
the survival of weight at energies of the order of the original spin-gap,
in regions where susceptibility
measurements show that the gap has collapsed (which occurs rapidly
upon doping according to the intuitive picture discussed 
in previous sections and before in Ref.\cite{sandvik}). Recently,
related work in the context of Zn-doped ladders have appeared with
conclusions similar to ours.\cite{fukuyama,motome,sigrist}

An important issue that we have not addressed in detail in this work is
the
behavior of the spin-spin correlations after Zn doping of ${\rm CuGeO_3}$.
If these correlations are $enhanced$ upon such doping, then a weak 
interchain coupling may lead to the stabilization of
a 3D N\'eel order as observed experimentally.

\begin{figure}[htbp]
\centerline{\psfig{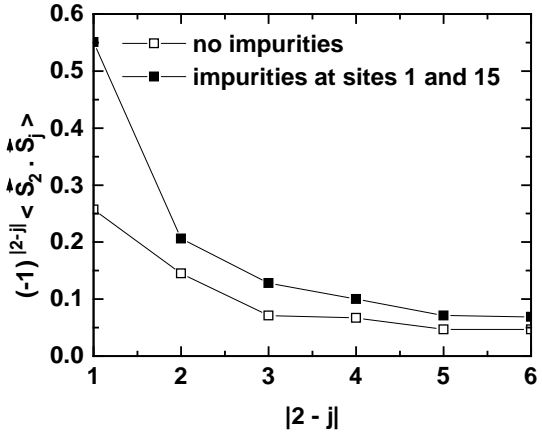}}
\vspace{0.5cm}
\caption{Spin-spin correlation as a function of distance for a ring of
16 sites with and without impurities located at sites 1 and 15. The
parameters are $J_2/J_1=0.24$ and $\delta=0.03$. The correlations are
measured with respect to site 2, i.e. the site immediately next to
an impurity.}
\label{fig8}
\end{figure}

Actually such an unusual effect (correlation
enhancement by Zn doping)
has been observed in a numerical study of  
ladders\cite{motome} and we have also detected such an enhancement in
one dimensional spin 1/2 chains as those studied here. For example, in
Fig.8 we show the spin-spin correlation measured on a ring of 16 sites
with impurities located at sites 1 and 15, compared with the
correlations
obtained on a ring with the same number of sites but without the
impurities. A remarkable enhancement of the correlations is observed
at short distances. 
Currently we are investigating the origin of this
enhancement which seems to be generic of a variety of Zn-doped materials.
Other systems with spin-gaps should behave similarly. S=1 chains have been
recently studied\cite{nio} and weight inside the Haldane gap was
reported. In two dimensions, a similar behavior should be
observed in Zn-doped cuprates in  the
temperature and density regime where a spin-gap 
is observed for the undoped case.\cite{sandvik} 

\section{acknowledgments}

We thank A. Sandvik, M. A. Continentino, S. Haas and A. Moreo
for fruitful discussions.
GBM acknowledges the financial support of the Conselho
Nacional de Desenvolvimento Cient\'{i}fico e Tecnol\'{o}gico (CNPq-Brazil).
ED is supported by grant NSF-DMR-9520776. Additional 
support by the National High
Magnetic Field Lab and the Supercomputer
Computations Research Institute at Florida State University
are acknowledged.

\end{document}